\documentclass[conference, a4paper, 9pt]{ieeeconf}
\IEEEoverridecommandlockouts     

\usepackage{nopageno}
\usepackage{soul}  
\usepackage{times}
\usepackage{array} 
\usepackage{helvet}
\usepackage{courier}
\usepackage{graphicx}
\usepackage{hyperref}
\usepackage{booktabs} 
\usepackage{multirow}
\usepackage{colortbl}
\usepackage{subcaption}
\usepackage{algorithm}
\usepackage{algpseudocode}
\usepackage{amsmath}
\usepackage{amsfonts}
\usepackage{soul}
\usepackage{xcolor}
\usepackage{authblk}
\usepackage[utf8]{inputenc}
\usepackage[T1]{fontenc}

\begin{document}

\title{\LARGE \bf Inferring Accurate Bus Trajectories from Noisy Estimated Arrival Time Records}

\author{Lakmal Meegahapola$^{1,a}$ \thanks{$^1$Idiap Research Institute \& École polytechnique fédérale de Lausanne (EPFL), Switzerland. email: lakmal.meegahapola@epfl.ch} 
\and 
Noel Athaide$^{2,a}$\thanks{$^{2}$InfoSys Limited, India}
\and
Kasthuri Jayarajah$^3$\thanks{$^3$Singapore Management University, Singapore}
\and Shili Xiang$^4$\thanks{$^4$Institute for Infocomm Research, Singapore}
\and
Archan Misra$^3$\thanks{$^a$Work done while being a Research Engineer at Singapore Management University, Singapore}
}


\maketitle
\thispagestyle{empty}
\pagestyle{empty}

\begin{abstract}
Urban commuting data has long been a vital source of understanding population mobility behaviour and has been widely adopted for various applications such as transport infrastructure planning and urban anomaly detection. While individual-specific transaction records (such as smart card (tap-in, tap-out) data or taxi trip records) hold a wealth of information, these are often \emph{private} data available only to the service provider (e.g., taxicab operator). In this work, we explore the utility in harnessing \emph{publicly available}, albeit noisy, transportation datasets, such as noisy ``Estimated Time of Arrival" (ETA) records (commonly available to commuters through transit Apps or electronic signages). We first propose a framework to extract accurate individual bus trajectories from such ETA records, and present results from both a primary city (Singapore) and a secondary city (London) to validate the techniques. Finally, we quantify the upper bound on the spatiotemporal resolution, of the reconstructed trajectory outputs, achieved by our proposed technique.
\end{abstract}

\section{Introduction}

In recent years, there has been an increased interest in the
use of individualised traces of mobility data for both transportation-related analytics and broader understanding of urban human behaviour--such data includes those obtained via 
GPS/WiFi sensing~\cite{Zhou16}, records of taxi ridership~\cite{Chiang15,jayarajah2018} or bike trips~\cite{Zhang18}, or consumer interactions with public transport~\cite{Yuan13, Meegahapola2019}). While smart card-based consumer transaction data can offer fine-grained insights (e.g., in detecting city-scale events~\cite{jayarajah2015event}), the reality is that access to such data is very difficult to obtain (often, due to privacy concerns). In many cases, such as (origin, destination) records of taxi trips or transaction data of ride sharing services, such data are in private hands and not universally available. 
Moreover, even for data based on public, governmental services, public authorities usually release such data with a significant delay. This implies that such individualised data may be useful for offline, longer-term policy or operational insights, but not always usable for applications requiring soft real-time responsiveness.


To overcome both these limitations, in this work, we exploit publicly available and \emph{aggregated} bus data, available as noisy Estimated Time of Arrivals (ETA), to tackle the problem of \emph{accurately inferring the transit times of individual buses at bus stops in soft-real time}. The ETA estimates, which contain no information about individual-specific trips but often include aggregate occupancy information, are often made available continuously via `live' public portals and well-document APIs, enabling external organisations to incorporate such data in third-party services. We believe that such data, when subject to intelligent processing, can  be a powerful information source for studying urban mobility.

We illustrate a scenario to motivate the use of such individual bus trajectory tracing: accurate estimates of the times at which a bus stops at different bus stops allows us to more precisely extract the occupancy level changes \emph{at} such stops. With longitudinal observations of such fine-grained (i.e., bus stop level) occupancy (or \emph{loading}) level changes, it is then possible to (a) build models for  \emph{normal} occupancy level changes, accounting for factors such as day of the week, time of the day, locality, etc., and (b) subsequently, use observed deviations from such normal patterns to not just detect anomalous events (e.g.,~\cite{jayarajah2015event, jayarajah2018}) but even anticipate or predict anomalies in advance (e.g., see~\cite{Meegahapola2019}).


To enable such diverse applications, we present and evaluate algorithms to accurately reconstruct the travel trajectory of vehicles from such publicly available ETA data. We make the following \emph{Key Contributions}:
\begin{enumerate}
\item{We propose an algorithmic framework to extract trajectories of individual buses from noisy ETA records even when the records do not contain individual bus-level identifiers. More specifically, the records periodically provide snapshot estimates of the ETA of the next $K$ ($K$ being a parameter) buses on a particular route, along with (optionally) the instantaneous GPS-based locations of the corresponding buses. The proposed approach first uses de-duplication to create a trajectory of a single bus instance across multiple snapshots, and then interpolates the trajectory data to infer the transit times at missing bus stops. Using a real-world dataset from Singapore, we evaluate the accuracy of the algorithms extensively and show that we can detect the \emph{exact} time a bus transits a downstream bus stop with more than 50\% precision/recall. More importantly, the precision/recall values exceed $80\%$, when we allow an estimation error of $\le$ 1 minute. We also show that the \emph{interpolation} step is crucial, as it decreases the error in estimating the arrival time of buses at downstream bus stops, by as much as $\approx$ 30\% for select bus services in Singapore. We also analyse how various other factors, such as (a) the day of the week (weekdays vs. weekend) and (b) the type of route (whether the bus goes through the city centre), affect the estimation accuracy.}

\item{By extending our analyses to data curated from a different city (i.e., London), we show that the performance is comparable (e.g., 40-50\% precision/recall in detecting arrivals with zero temporal error) and establish the generalisability of our technique. However, we also point out key differences: for example, unlike Singapore, specifying an estimation tolerance error of $\pm$1 minute does not significantly improve the precision/recall results for London.}

\item{Finally, we evaluate the impact of several practical choices on the achievable accuracy. First, by varying the frequency with which the ETA updates are generated, we show that more frequent updates can result in a significant (at least 20\% improvement in bus arrival time estimation accuracy). We additionally translate the impact of margins of time errors to practical upper bounds on the spatial resolution for different \emph{target} accuracy of bus arrival detection values; we show that for both Singapore and London, a bus arrival time detection accuracy of $\approx$ 80\% can be reached with a nominal spatial error (i.e., distance from the bus stop) of 200-600 meters. }
\end{enumerate} 

\section{Overall Methodology}
\label{sec:method}

\begin{figure*}[t]
    \centering
	\includegraphics[width=0.9\textwidth]{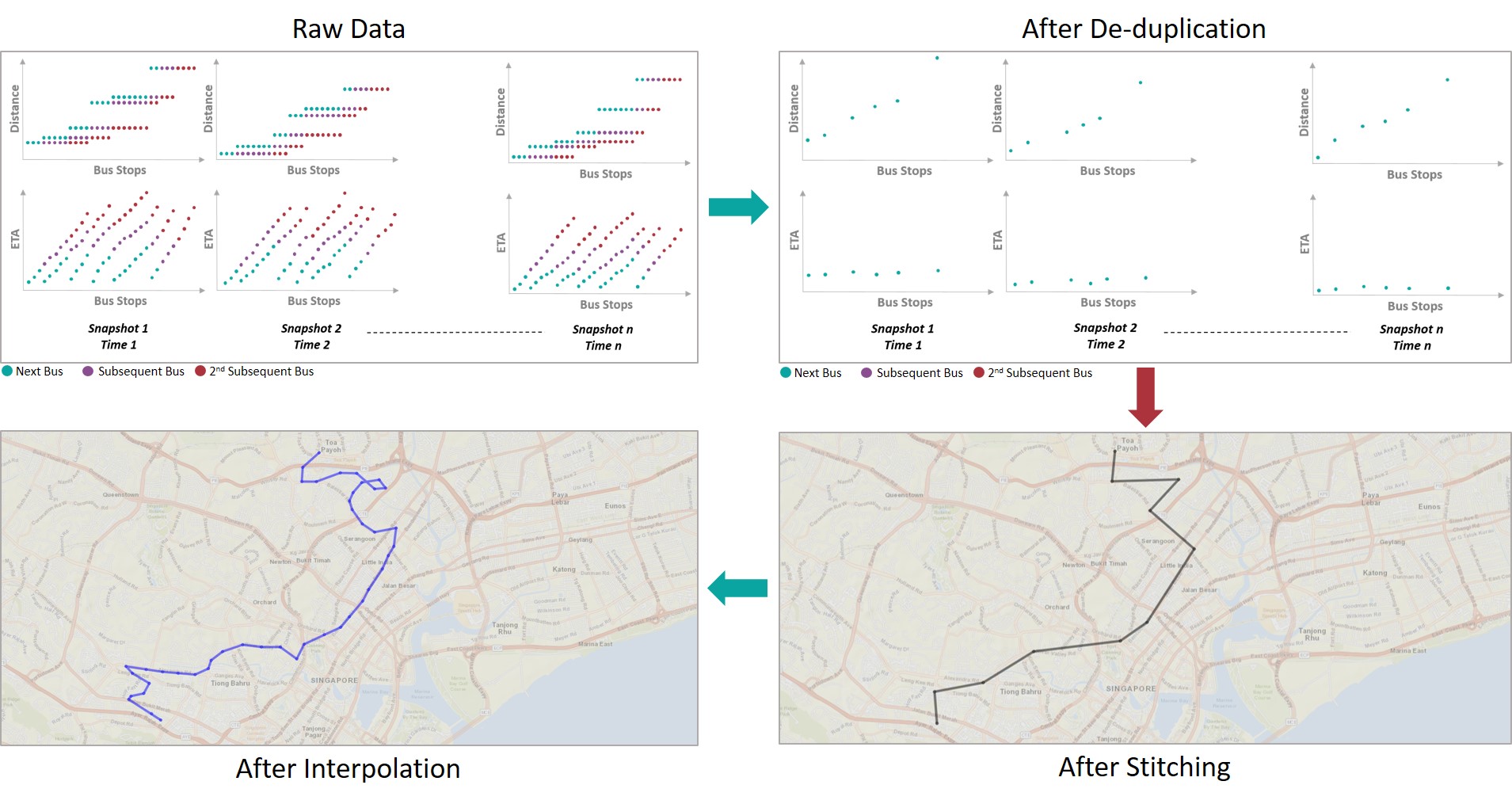}
\caption{Overview of the algorithms described in this work.}
\vspace{-0.2in}
\label{fig:framework}
\end{figure*}

We consider a dataset D consisting of tuples: $<date, t, busstopID, service, eta, latitude, longitude, K>$. Here, $date$ and $t$ corresponds to the calendar day and the discretized time interval the observation belongs to, $busstopID$ and $service$ identify a unique service route and a bus stop on that route, the $eta$ denotes the estimated time of arrival (in minutes) of the next $K^{th}$ bus, and the (optional) $latitude, longitude$ pair captures the bus' current instantaneous location. We compute the $distance$ a bus has traveled since the starting stop based on the bus' current location. 

\textbf{Problem Statement: }Given a sequence of bus stops, $S_{n}, n \in [1, N]$,  for a given $service$ with $N$ stops, and $D$, find $m$ unique buses over the course of the $day$ and their respective trajectories, $T_m$, which is a sequence of tuples $(m, n, t_{arrival})$  where $t_{arrival}$ is the time at which a bus $m$ arrives at stop $S_{n}$ along route $service$.

To this end, we describe our framework that consumes raw ETA observations to extract out unique trajectories of individual buses, as follows:
\begin{itemize}
\item{\textbf{Step 1: De-duplication -- }}In this step, we consider pairs of consecutive bus stops to remove duplicate observations of the same bus and assign \emph{local} identities within the same time instance (see Section~\ref{sec:deduplication}).
\item{\textbf{Step 2: Stitching -- }Next, we consider such unique bus instances across consecutive snapshots in time, to extract the \emph{trajectory}, $T_m$, of that instance (see Section~\ref{sec:stitching}).}
\item{\textbf{Step 3: Interpolation --} Finally, we perform interpolation to \emph{fill} the resulting trajectory with missed (due to removal of duplicates from Step 1) transits (see Section~\ref{sec:interpolation}).}
\end{itemize}

\subsection{Snapshot Deduplication}
\label{sec:deduplication}
We define a \textbf{snapshot}, $SS_{i, service} \subset D$, as the collection of all such tuples at $t=i$ for a representative $service$. Figure~\ref{fig:framework}a illustrates the $distance$ (top) and $eta$ (bottom) of multiple, consecutive \emph{snapshots} of ETA observations for a representative service across all its bus stops (on the $x-$axis). The illustration shows instances of the same physical $K^{th}$ bus across multiple consecutive bus stops as those instances where a bus has \emph{traveled} the same $distance$ since the start of the journey (points marked in the same color in the illustration the same physical bus) within each snapshot of time. In the deduplication stage, we eliminate such duplicate instances of the same bus, as described in Algorithm~\ref{alg:deduplication}, and illustrated in Figure~\ref{fig:framework}b.

In the deduplication stage, we eliminate such duplicate instances of the same bus, as described in Algorithm~\ref{alg:deduplication}, and illustrated in Figure~\ref{fig:framework}b. We limit ourselves to $K=3$ as the datasets used in this work (see Section~\ref{sec:data}) only provide information as advanced as the subsequent 3 buses. For two consecutive bus stops ($S_{n}$ and $S_{n+1}$), we consider the distance vectors of incoming $K=3$ buses, $X_{n}$ and $X_{n+1}$, respectively. We then find the optimal alignment, or 1-to-1 mapping, between the distance vectors. In this work, we resort to a simple Euclidean distance-based cost computation (lines 9 to 16) due to the number of observations/points being limited to $K=3$. This results in three cases: (1) the three buses seen by downstream bus stop $S_{n+1}$, are the same seen by $S_{n}$ during the same snapshot (lines 9 to 10), (2) $S_{n+1}$'s $K=1$ bus has already \emph{crossed} $S_{n}$ (lines 11 to 13), (3) $S_{n+1}$'s $K=1, 2$ buses have already \emph{crossed} $S_{n}$ (lines 14 to 16). After bus IDs are assigned \emph{locally}, between pairs of consecutive bus stops, the duplicate instances of a single bus are \emph{discarded} from the downstream bus stop. We refer to this as, \texttt{dist},  the distance-based deduplication technique. 

In Section~\ref{sec:eval}, we compare this against \texttt{ETA}-based de-duplication where the distance vectors are replaced by ETA-vectors. We consider an additional baseline, parametric-distance based de-duplication, which we refer to as \texttt{pdist}; here, we simply group together observations that are within $\epsilon$ distance of each other.  As, too small a value of $\epsilon$ may result in duplicate observations being treated as distinct, and too large a value of $\epsilon$ will result in distinct observations being treated as duplicates, we resort to finding an optimal value empirically. We varied the threshold between 0 to 4000 meters, and chose the $\epsilon$ that resulted in the lowest number of snapshots that failed a success criteria. In our experiments, we set this criteria to be an ETA between two consecutive bus stops to be no more than 10 minutes, across all bus stops. In contrast to \texttt{dist} and \texttt{eta}, \texttt{pdist} works in an \emph{offline} fashion as it relies on empirically learning a distance parameter that requires complete observations of a bus' journey between its terminal stops. As the applications considered in this work are geared towards \emph{online}, soft real-time response times, \texttt{dist} and \texttt{eta} based de-duplication are more appropriate, but are evaluated against \texttt{pdist} for accuracy comparisons.

\begin{algorithm}
\caption{Local Bus ID Assignment for Stop $S_{n+1}$ given bus IDs for observations at stop $S_{n}$}
\label{alg:deduplication}
\begin{algorithmic}[1]
\State \textbf{Inputs: } 
\State $X_n \leftarrow $ vector of bus distances observed at stop $S_n$ sorted desc.
\State $X_{n+1} \leftarrow $ vector of bus distances observed at stop $S_{n+1}$ sorted desc.
\State $I_n \leftarrow $ vector of IDs for buses observed at stop $S_n$ 

\State \textbf{Output: $I_{n+1} \leftarrow$ vector of IDs for buses observed at stop $S_{n+1}$}  

\State $cost\_rr\_gg\_bb = ((X_n[1] - X_{n+1}[1]) + (X_n[2] - X_{n+1}[2]) + (X_n[3] - X_{n+1}[3]))/3$
\State $cost\_rg\_gb = ((X_n[1] - X_{n+1}[2]) + (X_n[2] - X_{n+1}[3]))/2$
\State $cost\_rb = (X_n[1] - X_{n+1}[3])$

\If{$cost\_rr\_gg\_bb$ is least} 
\State $I_{n+1} = I_n$

\ElsIf{$cost\_rg\_gb$ is least }
\Comment{Generate/Increment new ID} 
\State $busID = I_n[maxLength] + 1 $
\Comment{Append new bus to stop $S_{n+1}$}
\State $I_{n+1} = append(I_n, busID)$

\ElsIf{$cost\_rg\_gb$ is least }
\Comment{Generate/Increment new IDs} 
\State $busIDs = I_n[maxLength] + 1, I_n[maxLength] + 2 $
\Comment{Append new buses to stop $S_{n+1}$}
\State $I_{n+1} = append(I_n, busIDs)$
\EndIf
\end{algorithmic}
\end{algorithm}

\subsection{Trajectory Stitching}
\label{sec:stitching}
In the previous step, we match buses across pairs of bus stops within the same time snapshot. In this step, we take the resulting bus ID assignments, and match bus IDs across time, i.e., \emph{snapshots}, to complete the traces of individual buses, as outlined in Algorithm~\ref{alg:stitching}. Figure~\ref{fig:framework}c illustrates the \emph{stitched trajectory} of a single bus (spanning multiple bus stops along its route and multiple snapshots).

\begin{figure}[t]
    \centering
	\includegraphics[width=\linewidth]{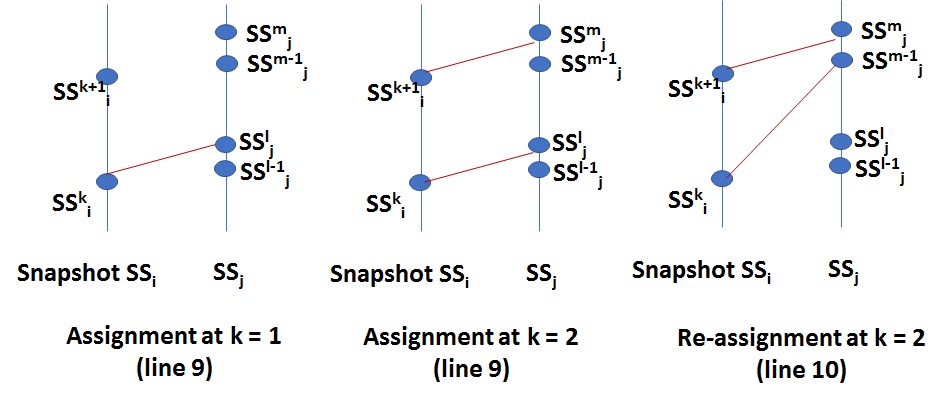}
\caption{Illustration of Lines 9 and 10 of Algorithm 2.}
\vspace{-0.15in}
\label{fig:algorithm2}
\end{figure}

Similar to the previous step, we rely on simple distance-based measurements for matching, and introduce new buses where a suitable match is not found (lines 6 to 16). We illustrate the workings of Lines 9 and 10 in Figure~\ref{fig:algorithm2} for clarity. Additionally, we introduce a speed criteria as a means for sanity check (lines 17 to 23);  the speed of each bus is computed given a successful match, and if the speed of any bus is greater than some threshold (in our case, we set it to 40kmph four our experiments) then we clear that matching, skip the current snapshot and repeat the matching process. The astute reader will note that due to the removal of duplicate bus instances from the previous step, the trajectory output here from this step will be a proper subset $S_{n}' \subset S_{n}$. 

\begin{algorithm}
\caption{Assign bus IDs to observations in snapshot $SS_j$ given bus IDs for observations in snapshot $SS_i$}
\label{alg:stitching}
\begin{algorithmic}[1]
\State $i = 1$ 
\State $j = 2$ 
\While{j <= no. of snapshots}
    \State $SS_i \leftarrow $ $i^{th}$ snapshot 
    \State $SS_j \leftarrow $ $j^{th}$ snapshot 
    \For{$k = 1$ to no. of observations in $SS_i$}
        \State $C \leftarrow$ set of observations from $SS_j$ between $(SS^k_i, SS^{k+1}_i)$
        \If{$C$ is not empty}
            \State $l \leftarrow argmax_{l \in |C|}{dist(SS^k_i, SS^l_j)}$
            \State move up RHS of all previous matchings by $|C|-1$
            \State if any unmatched observations left from 1 to $l-1$ in $SS_j$, assign new bus IDs to them 
        \Else
            \State match $SS^k_i$ to nearest neighbor in $SS_j$ from $SS^{k+1}_i$ which has travelled a greater distance
        \EndIf
    \EndFor
    \If{above matching violates speed check}
        \State clear matching and $j \leftarrow j + 1$
        \Comment{skip to next snapshot}
    \Else
        \State $i \leftarrow j$, $j \leftarrow i + 1$
        \Comment{Save matching}
        \State 
    \EndIf
\EndWhile
\end{algorithmic}
\end{algorithm}

\subsection{Interpolation}
\label{sec:interpolation}
In the final step, we fill in for \emph{missed} tuples (corresponding to missed bus stops due to the de-duplication step from before) in $T_m$. To this end, for each missed stop ($S_{missed}$) along the route, we consider the (1) estimated time of arrival $t_{arrival}^{missed-1}$ at the previous stop ($S_{missed-1}$) in the current trace,  (2) the distance between the pair of stops, $d$, and (3) the historical average velocity of any bus along that service, $v$, and estimate the time of arrival, $t_{arrival}^{missed}$ as $t_{arrival}^{missed-1} + \frac{d}{v}$. In Figure~\ref{fig:framework}d, we observe that the interpolate trajectory now consists of a more accurate representation of the bus (transiting through every stop along its route) in comparison to the stitched trajectory from before. Now, $\vert T_m \vert = N$.

\section{Data Description}
\label{sec:data}
We consider a longitudinal, bus dataset from Singapore, as our primary dataset in this work. We collected Estimated Arrival Time (ETA) records of incoming buses for over 401 bus services from over 4913 bus stops island-wide using the publicly available DataMall API \footnote{\url{https://www.mytransport.sg/content/mytransport/home/dataMall.html}}. Each record is a tuple $<busstopid, serviceid, timestamp, loading, ETA, loc_x, loc_y>$ where $busstopid$ and $serviceid$ identify the bus stop and the service, respectively, and $loading$ is a discrete number varying between 1 and 3 capturing the level of crowdedness of the next bus scheduled to arrive at that stop, the $ETA$ representing the estimated arrival time of the bus, and $loc_x$ and $loc_y$ representing the GPS coordinates in at query time. The API is refreshed every 60 seconds, with $timestamp$ capturing the most recent refresh instant. In addition to the next immediate incoming bus, the data also provides the same information of the subsequent bus and the bus thereafter, when available. We consider the locations of bus stops and the service route information \footnote{https://www.mytransport.sg/content/mytransport/home/dataMall/dynamic-data.html} to convert the two-dimensional location of incoming buses to a uni-dimensional \emph{distance} measure, i.e., the distance the bus has traversed since its origin stop. 

Additionally, we collected data from London, using the Transport for London Unified API \footnote{https://tfl.gov.uk/info-for/open-data-users/unified-api}, to validate the methodology presented in this work. In contrast to the Singapore dataset, the presence of a field capturing the identity of the vehicle allows us to use this dataset as means for additional validation (see Section~\ref{sec:eval}) albeit not consisting of the GPS coordinates of the buses' instantaneous location. The records are refreshed every 30 seconds. We summarize key statistics related to the datasets in Table~\ref{tab:data}.

\begin{table}[!tbh]
\vspace{-0.1in}
  \begin{tabular}{ccccc}
    \toprule
    City & Observation Period & Services Considered & Sampling Rate \\
    \midrule
    Singapore & 2018-06-01 to & 124, 147, 190 & 60 seconds\\
     & 2018-07-19 & 139, 32, 65, 195 & \\
    London  & 2018-07-04 to 08 & 12, 17, 326, 111 & 30 seconds\\\hline
    \bottomrule
  \end{tabular}
  \caption{Summary of datasets used in the analysis. }
  \vspace{-0.3in}
  \label{tab:data}
  
\end{table}

\subsection{Preliminaries}
As we mentioned previously, our methods for extracting bus trajectories from ETA records are inherently noisier, as compared to, the often \emph{protected} smart fare card transactions-based data due to: (1) the inability to distinguish between individual buses (as the bus IDs are usually unknown), (2) and the timing of when a bus actually crosses the individual stops along its route being unknown. 

Here, we discuss some key scenarios (that we observe in our dataset) that challenge accurate trajectory discovery. To quantify the intensity of these possible problem scenarios, we rely on an additional dataset consisting of smart-fare based transactions which consist of the details of trips made by commuters, including start and end stops, start and end time of the trip, the service route and registration number of the bus, during the month of August 2013.

\textbf{Paired trajectories: }We observe a non-negligible number of instances (e.g., 43.96\% cases over in service 190 \& 22.94\% cases over 5 service routes) where two or more buses servicing the same route to be closely following each other, especially during peak hours. Without knowing the identity of the individual buses, the algorithms presented in this work are likely to cluster such buses as a single bus instance.

\textbf{Skipped stops: }As commonly seen across cities, buses skip certain bus stops when there are no alighting or boarding passengers for an upcoming stop. We observe that on a typical weekday, on average  24.78\% stops are skipped. 

\begin{figure}[!tbp]
\vspace{0.1in}
  \centering
    \includegraphics[width=0.6\linewidth]{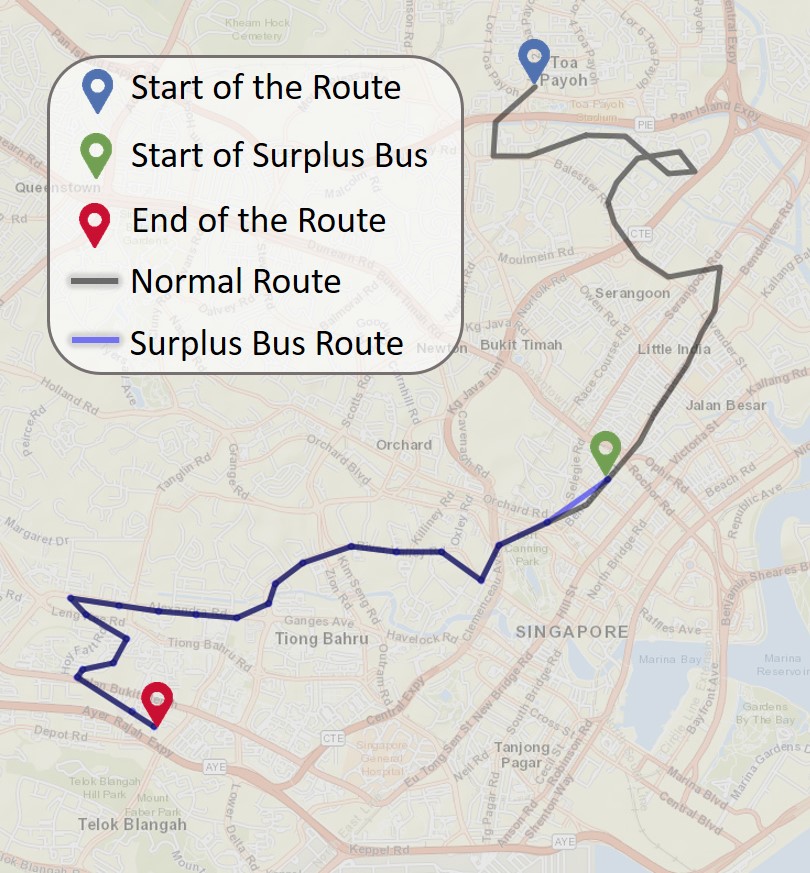}
    \caption{Original Route and the Route of a Surplus Bus in Route 139, Direction 1}
    \captionsetup{justification=centering}
    \label{fig:surplus-bus-route}
\end{figure}

\textbf{On-demand activation of surplus buses: }Typically, buses start and end their journeys at stipulated terminus stops for that service route. In our data, we observe that, especially during peak hours, there are a number of instances (e.g., 5.29\% additional buses in operation, on a typical weekday PM peak hours) where buses are activated \emph{on-demand}, which don't necessarily start their journey from the beginning of the route (illustrated in Figure~\ref{fig:surplus-bus-route}) which introduces errors as the algorithms take into account, the distance the bus has travelled since the start.

\begin{figure*}[t]
\centering
        \begin{subfigure}[t]{0.4\textwidth}
        \centering
	    \includegraphics[width=2.2in]{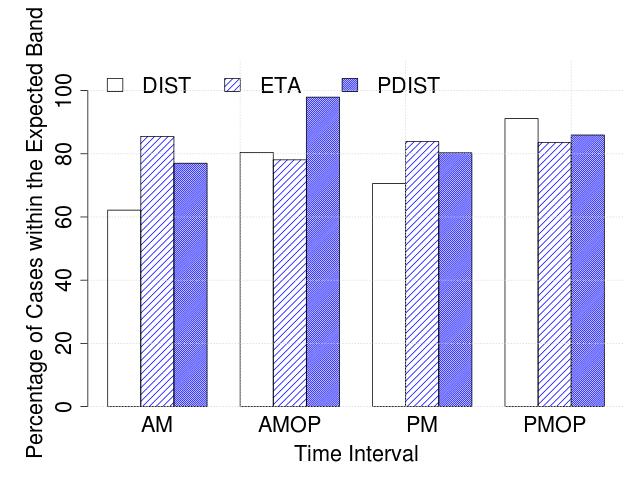} 
        \end{subfigure} 
        \begin{subfigure}[t]{0.4\textwidth}
        \centering
	    \includegraphics[width=2.2in]{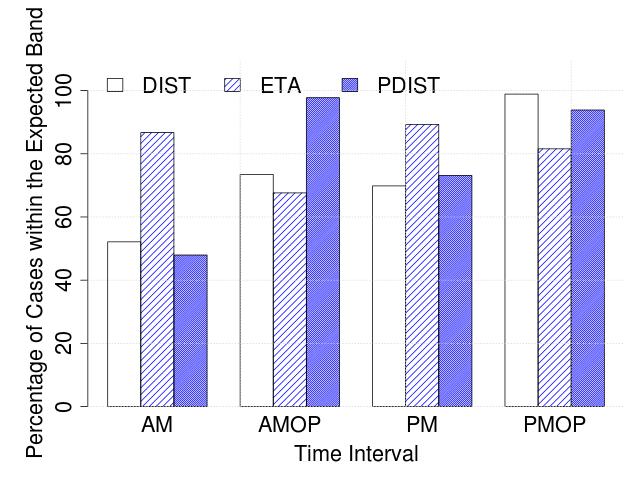} 
        \end{subfigure} 
\caption{Percentage of bus arrival estimation within the expected scheduled frequency, for (a) weekdays and (b) weekends, for different hours of the day. }
 \vspace{-0.2in}
\label{fig:freq}
\end{figure*}

\section{Experiments}
\label{sec:eval}
In this section, we summarize results from our experiments in evaluating our algorithms along a number of dimensions.

\subsection{Performance Metrics}
We evaluate three main aspects of our combined framework: (1) the accuracy of detecting individual buses, (2) the accuracy with which the crossing of a bus at a particular bus stop is detected (i.e., arrival detection), in Section~\ref{sec:perf}, and (3) the impact of practical considerations such as sampling frequency on the limits to performance in Section~\ref{sec:practical}. 

We measure the effectiveness of the algorithm in detecting individual buses by comparing the frequency of detected buses against actual transit schedules. We report the percentage of cases where the algorithm under, or, over-reports the frequency for each $bus service$. 

For each distinct $<bus service, bus stop>$ pair, we consider buses that cross the stop as the set of \emph{observed} buses, and consider that the algorithm \emph{detected} that bus, if the arrival time of a bus (of the same service for that bus stop) is within a time, $\Delta_{t}$, of the actual arrival time $t_{actual}$ of that bus. We then follow standard definitions in reporting the accuracy in terms of precision (Eq. \ref{eq:precision}) and recall (Eq. \ref{eq:recall}) where $Obs$ is the set of all observed buses, $Det$ is the set of all detected buses, and $\left\vert.\right\vert$ denotes the cardinality operator.

\begin{equation}
\label{eq:precision}
precision = \frac{\left\vert Obs \cap Det \right\vert}{\left\vert Det \right\vert}
\end{equation} 

\begin{equation}
\label{eq:recall}
recall = \frac{\left\vert Obs \cap Det \right\vert}{\left\vert Obs \right\vert}
\end{equation} 


\subsection{Bus Trajectory Tracking}
\label{sec:perf}

\subsubsection{Frequency Analysis}
To understand the effectiveness of our algorithms, we first study at an aggregate frequency level over multiple time windows across the day. We consult publicly available schedule information\footnote{\url{https://www.transitlink.com.sg/eservice/eguide/service_idx.php}}, to extract the approximate number of buses expected to be in service for four different times of the day: AM (6-10:29am), AMOP (10.30am-3:59pm), PM (4pm-7.59pm), and PMOP (8pm-5:59am). In Figure~\ref{fig:freq}, we plot the percentage of cases where the detected number of buses by our algorithms (distance-based, ETA-based and parametric distance-based) were within the band reported in the schedule information, for weekdays and weekends, separately. We observe that the ETA-based method performs consistently better than the methods (except for during AM Off-Peak hours), and that the accuracy is $\geq$ 80\% in general, except for AM Off-Peak hours during the weekend.

Next, in Table~\ref{tab:freqweekdays} and Table~\ref{tab:freqweekends}, we summarize the accuracy of detection, for a sample set of service routes (i.e., 139, 65, 32 and 195). Except for service 195 which is \emph{looped}, the remaining three services serve in 2 directions. The zero-error margin results correspond to cases where we consider that the detection was correct only if the number of buses detected fall exactly within the \emph{scheduled} band, and the 10\% error margin corresponds to cases where the detections are within a +/- 10\% of the scheduled band. We see that the algorithms are less accurate over weekends and during off peak hours.

\begin{table}[b]
\vspace{-0.15in}
\resizebox{\linewidth}{!}{
  \begin{tabular}{p{1cm} | >{\centering\arraybackslash}m{0.5cm}  >{\centering\arraybackslash}m{0.5cm} 
>{\centering\arraybackslash}m{0.5cm} 
>{\centering\arraybackslash}m{0.5cm} |
>{\centering\arraybackslash}m{0.5cm}  
>{\centering\arraybackslash}m{0.5cm} 
>{\centering\arraybackslash}m{0.5cm} 
>{\centering\arraybackslash}m{0.5cm}}
    \toprule
    \textbf{Service} & \multicolumn{4}{c}{\textbf{Zero error margin}} &  \multicolumn{4}{c}{\textbf{10\% error margin}}\\
     & AM & AMOP & PM & PMOP & AM & AMOP & PM & PMOP \\
    \midrule
    139-1 & 14.8 & 27.4 & 88.7 & 51.6 & 75.4 & 64.5 & 100  & 98.4 \\
139-2 & 3.2  & 71   & 98.4 & 91.9 & 14.5 & 98.4 & 100  & 98.4 \\
32-1  & 88.9 & 77.8 & 98.4 & 71.4 & 95.2 & 95.2 & 100  & 95.2 \\
32-2  & 46.8 & 24.2 & 100  & 61.3 & 88.7 & 71   & 100  & 96.8 \\
65-1  & 33.9 & 33.9 & 100  & 66.1 & 88.7 & 67.7 & 100  & 91.9 \\
65-2  & 75   & 40.6 & 98.4 & 93.8 & 93.8 & 84.4 & 100  & 100  \\
195   & 50   & 6.5  & 14.5 & 0    & 90.3 & 12.9 & 85.5 & 6.5 \\\hline
    \bottomrule
    
  \end{tabular}
  }
  \caption{Summary of accuracy (in percentage) in detecting individual buses, on \textbf{weekdays} using the \emph{ETA}-based algorithm.}
  \label{tab:freqweekdays}
\end{table} 

\begin{table}[h]
\vspace{-0.1in}
\resizebox{\linewidth}{!}{
  \begin{tabular}{p{1cm} | >{\centering\arraybackslash}m{0.5cm}  >{\centering\arraybackslash}m{0.5cm} 
>{\centering\arraybackslash}m{0.5cm} 
>{\centering\arraybackslash}m{0.5cm} |
>{\centering\arraybackslash}m{0.5cm}  
>{\centering\arraybackslash}m{0.5cm} 
>{\centering\arraybackslash}m{0.5cm} 
>{\centering\arraybackslash}m{0.5cm}}
    \toprule
    \textbf{Service} & \multicolumn{4}{c}{\textbf{Zero error margin}} &  \multicolumn{4}{c}{\textbf{10\% error margin}}\\
     & AM & AMOP & PM & PMOP & AM & AMOP & PM & PMOP \\
    \midrule
    139-1 & 0    & 20   & 60   & 40   & 4    & 68   & 92   & 92   \\
139-2 & 0    & 57.7 & 96.2 & 88.5 & 0    & 96.2 & 100  & 96.2 \\
32-1  & 72   & 76   & 96   & 60   & 100  & 96   & 96   & 88   \\
32-2  & 3.8  & 7.7  & 96.2 & 73.1 & 88.5 & 61.5 & 100  & 96.2 \\
65-1  & 42.3 & 30.8 & 96.2 & 19.2 & 88.5 & 57.7 & 100  & 84.6 \\
65-2  & 34.6 & 30.8 & 96.2 & 57.7 & 92.3 & 65.4 & 96.2 & 96.2 \\
195   & 88   & 8    & 44   & 12   & 100  & 44   & 100  & 72\\\hline
    \bottomrule
    
  \end{tabular}
  }
  \caption{Summary of accuracy in detecting individual buses, for different bus services, on \textbf{weekends} using the \emph{ETA}-based algorithm.}
  \label{tab:freqweekends}
  
\end{table}

\subsubsection{Arrival Time Detection}
Here we focus on the ability of the algorithms in detecting the exact instance an individual bus crosses a downstream bus stop along its service route.

\textbf{Groundtruth data:} To report the accuracy of bus arrival detection, in the case of Singapore, we collected additional data where we manually observed and annotated buses crossing particular bus stops along with the time, at the minute level, at which they crossed ($t_{actual}$). We collected data over 7 days between 5 PM and 7 PM (i.e., evening peak hours) from two different bus stops, one within the Central Business District (CBD), and another in the outskirts of the CBD area (non-CBD). In total, we collected 28 hours of annotated bus crossings, consisting of 3173 bus crossings ($\left\vert Obs \right\vert$). 

\begin{figure}[t]
\vspace{-0.2in}
\centering
        \begin{subfigure}[t]{0.5\linewidth}
	    \includegraphics[width=1.7in]{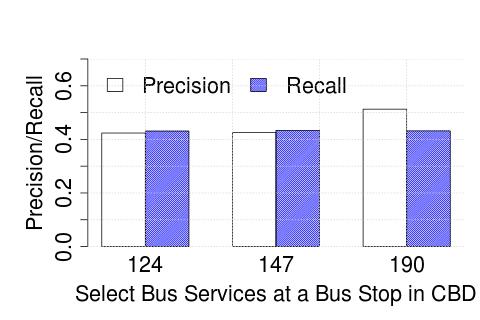} 
	    \caption{\texttt{DIST}, After De-duplication}
                  \label{fig:acc-dist}
        \end{subfigure}%
        \begin{subfigure}[t]{0.5\linewidth}
	    \includegraphics[width=1.7in]{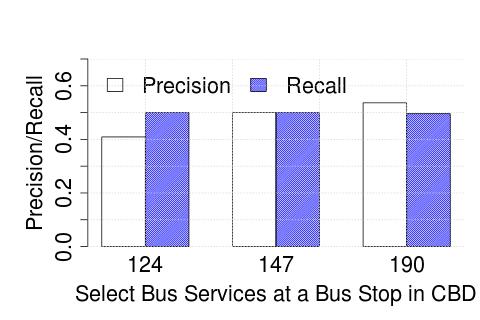} 
	    \caption{\texttt{ETA}, After De-duplication}
                  \label{fig:acc-eta}
        \end{subfigure}%
        \vfill
        \begin{subfigure}[t]{0.5\linewidth}
	    \includegraphics[width=1.7in]{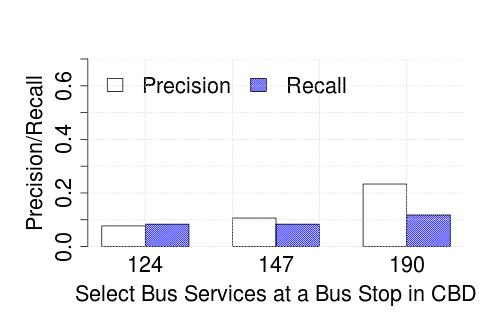} 
	    \caption{\texttt{ETA}, After Stitching}
                  \label{fig:acc-eta-stitched}
        \end{subfigure}%
        \begin{subfigure}[t]{0.5\linewidth}
	    \includegraphics[width=1.7  in]{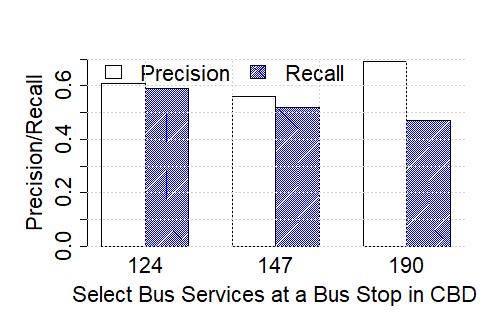} 
	    \caption{\texttt{ETA}, After Interpolation}
                  \label{fig:acc-eta-interp}
        \end{subfigure}%
\caption{Precision/Recall of Arrival Detection for (a) distance-based deduplication (\texttt{DIST}), (b) ETA-based deduplication (\texttt{ETA}), (c) ETA-based output after stitching, with $\Delta_t = 0$, and (d) after velocity-based interpolation, for Singapore.}
\vspace{-0.2in}
\label{fig:accuracy-sg}
\end{figure}

In Figure~\ref{fig:accuracy-sg}, we plot the precision and recall of detecting the exact time a given bus crossed a chosen bus stop (near our university campus), to the closest minute, for three different bus services, at different stages of the algorithm: de-duplication, stitching and interpolation. We find that the accuracy to be in the 40-50\% range while using \texttt{ETA} (while using distance alone performs generally worse). 

\begin{figure}[t]
\centering
        \begin{subfigure}[t]{0.5\linewidth}
	    \includegraphics[width=1.75in]{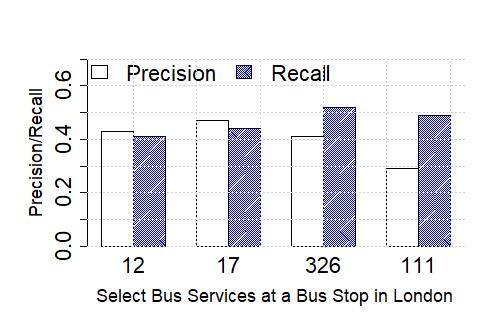}
	    \caption{\texttt{ETA}, After De-duplication}
                  \label{fig:accuracy-lon-dedup}
        \end{subfigure}%
        \begin{subfigure}[t]{0.5\linewidth}
	    \includegraphics[width=1.75in]{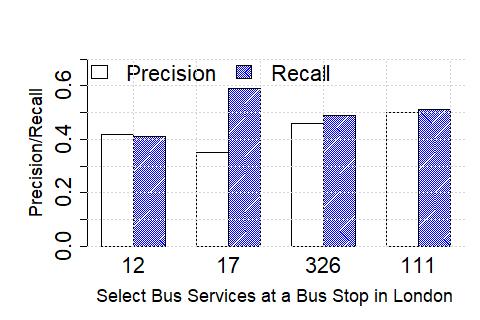} 
	    \caption{\texttt{ETA}, After Interpolation}
                  \label{fig:accuracy-lon-interp}
        \end{subfigure}%
\caption{Precision/Recall of Arrival Detection for (a) ETA-based de-duplication (\texttt{ETA}), and (b) after velocity-based interpolation, for London.}
\vspace{-0.2in}
\label{fig:accuracy-lon}
\end{figure}

As anticipated, the accuracy after the stitching stage is poor -- this is due to the fact that this stage, by design, removes multiple occurrences of a bus's trajectory. The interpolation stage then use these sparse, albeit accurate observations, to output the final trajectory. In Figure~\ref{fig:accuracy-lon}, we plot the accuracy values for the case of London, and note that the performance remains comparable.

\begin{figure}[t]
\centering
        \begin{subfigure}[t]{0.5\linewidth}
	    \includegraphics[width=1.8in]{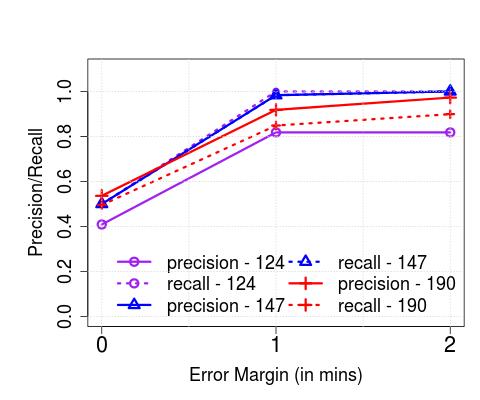} 
	    \caption{Singapore}
                  \label{fig:error-dedup-sg}
        \end{subfigure}%
        \hfill
        \begin{subfigure}[t]{0.5\linewidth}
	    \includegraphics[width=1.8in]{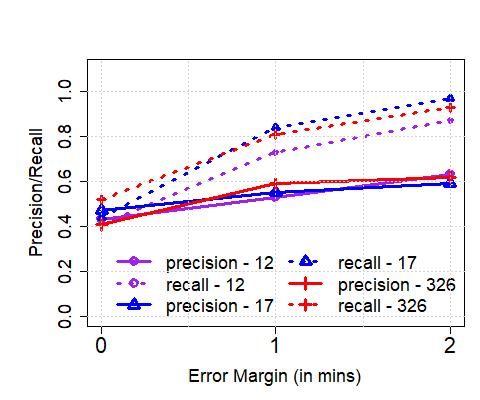}
	    \caption{London}
                  \label{fig:error-dedup-lon}
        \end{subfigure}%
\caption{Precision/Recall of Arrival Detection for varying time error margins ($\Delta_t$) for each bus service in the CBD after deduplication for (a) Singapore and (b) London.}
 \vspace{-0.2in}
\label{fig:accuracy-error-margin}
\end{figure}

\begin{figure}[t]
\centering
        \begin{subfigure}[t]{0.5\linewidth}
	    \includegraphics[width=1.8in]{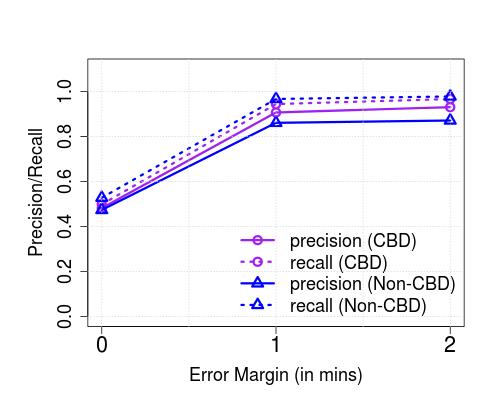} 
	    \caption{Sampled every 2 minutes (Singapore)}
                  \label{fig:error-coarse-sg}
        \end{subfigure}%
        \begin{subfigure}[t]{0.5\linewidth}
	    \includegraphics[width=1.8in]{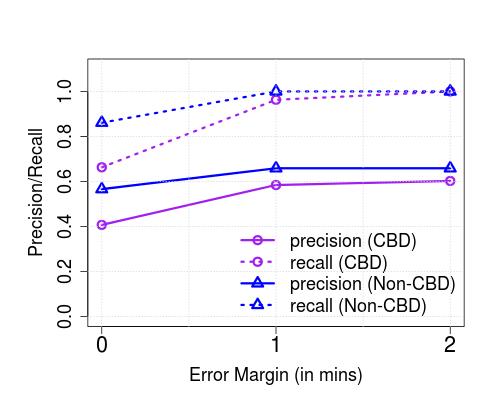} 
	    \caption{Sampled every 30 seconds (Singapore)}
                  \label{fig:error-fine-sg}
        \end{subfigure}%
        \vfill
        \begin{subfigure}[t]{0.5\linewidth}
	    \includegraphics[width=1.8in]{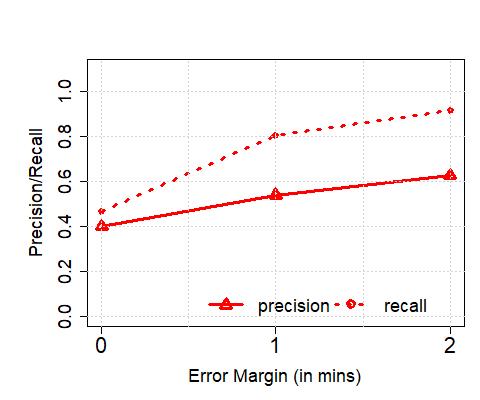}
	    \caption{Sampled every 2 minutes (London)}
                  \label{fig:error-coarse-lon}
        \end{subfigure}%
        \begin{subfigure}[t]{0.5\linewidth}
	    \includegraphics[width=1.8in]{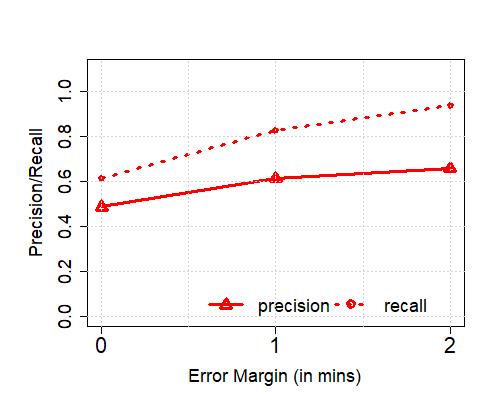}
	    \caption{Sampled every 30 seconds (London)}
                  \label{fig:error-fine-lon}
        \end{subfigure}%
\caption{Precision/Recall of Arrival Detection for varying time error margins ($\Delta_t$) with bus services averaged in CBD vs. non-CBD areas, in (after de-duplication) for different sampling frequencies, in Singapore and London.}
 \vspace{-0.2in}
\label{fig:accuracy-sampling}
\end{figure} 

\subsection{Practical Implications}
\label{sec:practical}
\textbf{Performance vs. Error Margin ($\Delta_t$): }Thus far, we have evaluated the effectiveness of our algorithms in distinguishing between distinct, physical buses and the ability to detect exactly when a bus crosses its downstream stops. However, we note that for several practical applications, knowing the \emph{exact} time of arrival (i.e., a zero margin of error) may not be critical. Here, we increase the error margin systematically, and observe its impact on the performance. In Figure~\ref{fig:accuracy-error-margin}, we plot the allowed error margin on the $x-$axis (varied between 0, 1 and 2 minutes), and the precision/recall on the $y-$axis for (a) Singapore and (b) London. We observe that the performance reaches its maximum with a minute's error, in both cases -- in other words, our algorithms are able to detect \emph{most} arrivals (with precision/recall $\geq 0.80$) within a minute of them occurring. We also observe that allowing for larger margins ($\geq 2$ mins) doesn't necessarily increase the performance.

\textbf{Performance vs. Sampling Frequency ($F_s$): }Next, we study the impact of the sampling frequency on the effectiveness of the algorithms. Note that thus far, we have resorted to $F_s = $2 minutes (\emph{low}) and $F_s = $30 seconds (\emph{high}), for Singapore and London, respectively, as were the recommended update frequencies. In Figure~\ref{fig:accuracy-sampling}, we plot the precision/recall of arrival detections, for varying permissible error margins, for both \emph{low} and \emph{high} frequencies, for services running through the CBD and otherwise. We observe that by more frequent polling, the recall improves significantly (e.g., from 50\% to 85\% in the case of non-CBD services in Singapore, with zero temporal error). However, we also note that frequent polling results in no observable improvement with increased error margins. We also point out, that whilst we see no significant difference in performance between CBD and non-CBD areas, with more frequent polling, we see that the algorithms perform better in detecting arrival times in the non-CBD area (e.g., $\approx$ 20\% improvement in both precision and recall with zero time error, over CBD buses).

\begin{figure}[t]
\centering
        \begin{subfigure}[b]{0.5\linewidth}
	    \includegraphics[width=1.8in, height=1.5in]{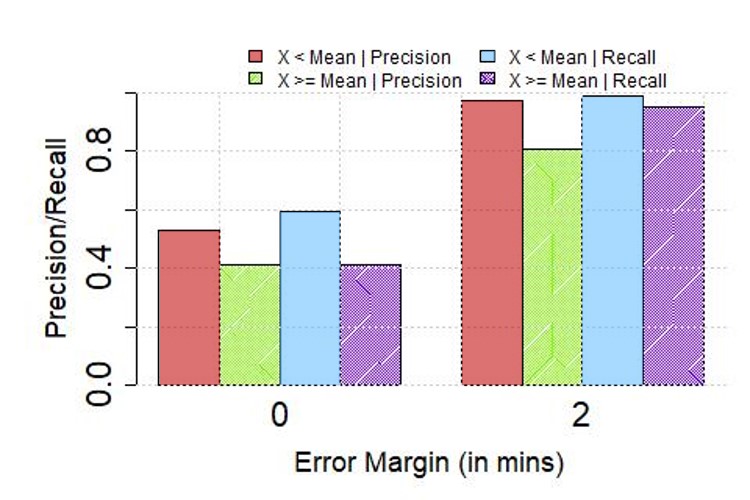} 
	    \caption{Singapore}
                  \label{fig:accuracy-interpolated-singapore}
        \end{subfigure}%
        \hfill
        \begin{subfigure}[b]{0.5\linewidth}
	    \includegraphics[width=1.8in, height=1.5in]{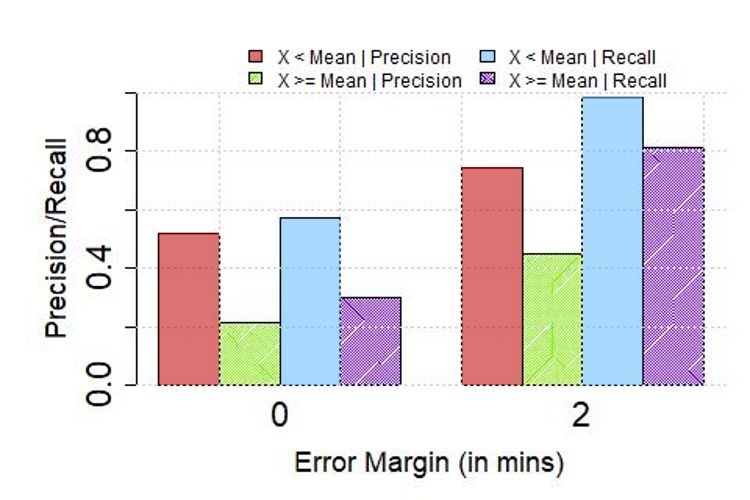}
	    \caption{London}
                  \label{fig:accuracy-interpolated-london}
        \end{subfigure}%
\caption{Performance stratified by distance in interpolation with sampling rate of 2 minutes.}
 \vspace{-0.3in}
\label{fig:accuracy-interpolated}
\end{figure}

\textbf{Performance vs. Error in Interpolation: }As we describe in Section~\ref{sec:method}, the last step in our algorithm, interpolation, outputs the arrival times of buses at each stop downstream when the corresponding data points are removed during the de-duplication stage. As this might introduce errors, we next study the extent to which such approximation errors impact the overall performance. In Figure~\ref{fig:accuracy-interpolated}, we plot the precision/recall for services running in CBD and non-CBD, for zero error margin, in (a) Singapore and (b) London, for \emph{low} (i.e., 1239.4 m or less in Singapore, 1032.7m or less in London -- these values are mean values for each city) and \emph{high} interpolation distance error. As anticipated, we note that the observed accuracy is lower for cases where we make a possibly larger error during interpolation (i.e., distance between the closest upstream and downstream bus stops of a stop of interest is large) -- for instance, we see that in the case of Singapore, with zero error margin, we see at least a 15-20\% increase in both precision and recall for cases where the interpolation error is \emph{lower} than the mean error over all cases. For London, we see that this improvement is more pronounced ($approx$ 30\%).

\begin{figure}[t]
\centering
        \begin{subfigure}[b]{0.45\linewidth}
	    \includegraphics[width=\textwidth]{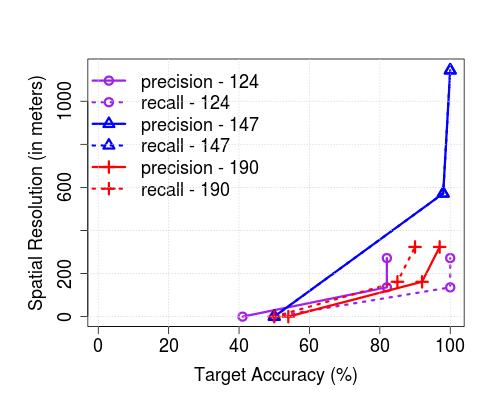} 
	    \caption{CBD}
                  \label{fig:spatial-dedup-cbd}
        \end{subfigure}%
        \begin{subfigure}[b]{0.45\linewidth}
	    \includegraphics[width=\textwidth]{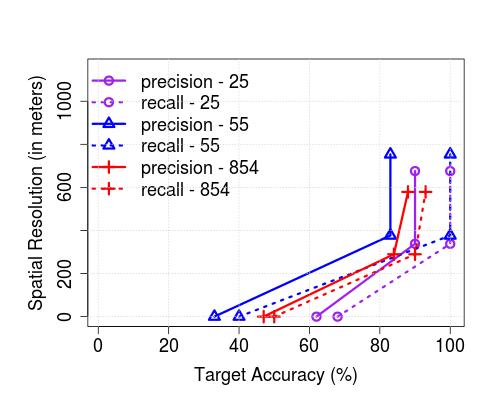} 
	    \caption{Non-CBD}
                  \label{fig:spatial-dedup-ncbd}
        \end{subfigure}%
\caption{Upper bound on spatial resolution for practical applications with de-duplication, for three bus services each in the (1) CBD area and (2) non-CBD area.}
 \vspace{-0.3in}
\label{fig:accuracy-spatial-dedup}
\end{figure}

\begin{figure}[!tbp]
  \centering
    \includegraphics[width=0.7\linewidth]{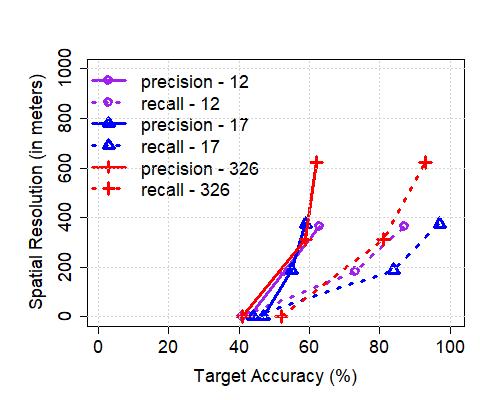}
    \caption{Upper bound on spatial resolution for practical applications after the de-duplication phase, for three bus services each in the London dataset.}
     \vspace{-0.3in}
    \label{fig:accuracy-spatial-lon}
\end{figure}

\textbf{Upper Bound on Spatial Resolution: }As a result of the \emph{temporal range} or \emph{error margins} within which the algorithms detect arrivals, there is an inherent upper bound on the \emph{spatial resolution} with which applications that consume such resulting bus trajectory data (coupled with occupancy levels) can operate practically. In order to understand this, we plot the achievable, or target accuracy, of a potential application, on the $x-$ axis, and the spatial resolution on the $y-$axis, after the de-duplication stage in Figure~\ref{fig:accuracy-spatial-dedup}. We compute the spatial resolution by first extracting the time error margin (extending beyond the maximum of 2 minutes we have thus far considered, where required) that results in the corresponding target accuracy (solid lines correspond to precision, and dashed lines correspond to recall), and then multiplying the temporal error by the average speed of any bus traversing through the corresponding service route. Unsurprisingly, the spatial resolution is significantly less for buses servicing non-CBD routes (e.g., maximum achievable accuracy is reached with a resolution $\leq$ 400 m in CBD (with the exception of precision for the 147 route) whilst the same is achieved with resolution in the 700 meters range in non-CBD areas). As an aside, we also note that the maximum achievable accuracy is service route-dependent and does not monotonically increase with increased permissible error margins (but saturates beyond a point, as observed previously). Further, in Figure~\ref{fig:accuracy-spatial-lon}, we observe that the spatial resolution achieved in London compare similarly -- for instance, for recalls of between 80-100\%, we observe that in both cases, Singapore and London, the resolution is in the 200-600 meters range. 




\subsection{An Illustrative Use Case: Bus Arrival Time Prediction}
\label{sec:usecase}
\begin{figure}[!tbp]
  \centering
    \includegraphics[width=0.6\linewidth]{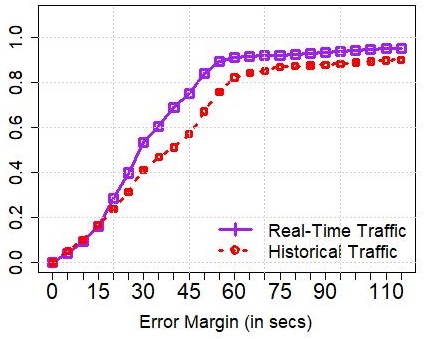}
    \caption{CDF of temporal error margins for bus arrival time predictions with \emph{real-time, stitched trajectories} vs. historical ETA records.}
    \label{fig:usecase}
\vspace{-0.15in}
\end{figure}

We anticipate that the ability to accurately trace the trajectory of public buses, \emph{at real time}, using publicly available \emph{ETA} records alone can be useful for many practical applications. We take an example application to demonstrate the utility of using real-time trajectory information of buses as opposed to relying on historic records -- in particular, in Figure~\ref{fig:usecase}, we plot observations from \emph{predicting} arrival times of buses for downstream bus stops.

For this analysis, we consider bus services \{7,14,16,36,147,190 and 857\} all of which pass through the bus stop ($S_{n}$) considered in Section~\ref{sec:eval} between the PM peak period of 5 PM and 7 PM, over all weekdays. In the real-time case, we consider observations an hour prior to the prediction time, to compute the average speeds of buses along the road segment that leads to this specific bus stop ($S_{n}$) from the upstream bus stop ($S_{n-1}$), depending on the service route). The prediction task involves predicting the \emph{estimated arrival time} of the outgoing buses at the next stop, during the test period of 5 PM to 7 PM on a specific day (i.e., June 6th 2018), based on the actual time of arrival at $S_{n-1}$, the average speed along the segment $S_{n-1} \rightarrow S_{n}$ and the length of the segment. The speed information here is computed on a per-bus basis due to the individual bus-level trajectory tracing made available by the technique described in the earlier sections.

In the historic case, we compute the speed along the segment based on the GPS traces of the unprocessed \emph{next buses} over weekdays from two past months (April to May 2018). In Figure~\ref{fig:usecase},the $x-$axis represents the temporal error margins over predictions for all buses during the two-hour test window whilst the $y-$axis represents the CDF for the two cases: (1) real-time, from stitched trajectories, vs. (2) historic, unstitched ETA records. We observe that while both methods are able to predict 20\% of the cases with a small error margin (e.g., $\leq$20 seconds), the real-time method outperforms for the remaining 80\% of the cases. For instance, whilst 75\% of the predictions are within an error margin of 45 seconds with the real-time data, the same is true only for 55\% of the cases using historic records. This observation clearly demonstrates the added benefit of being able to reconstruct trajectories at real-time.


\section{Related Work}
\label{sec:related}
Using transportation data for urban planning and analytics has a long and fruitful history (obtained via taxi ridership~\cite{Chiang15} or bike trip~\cite{Zhang18} records, public transport data~\cite{Yuan13, Meegahapola2019}). Recently, there have been a wealth of research that study traffic flow estimations using large datasets \cite{meng2017city, hoang2016fccf, li2017urban, bessa2016riobusdata}. While Meng et al. \cite{meng2017city} address the problem of estimating city-wide traffic volumes using loop detectors deployed across the city and taxi trajectories, Hoang et al.  \cite{hoang2016fccf} attempt to forecast city-wide crowd-flows based on big data. Li et al. \cite{li2017urban} demonstrate promise in predicting travel times using only a small number of \emph{GPS} cars. In other studies, researchers have also used transportation and mobility data to interesting and varied uses such as automatic transit map generation \cite{verma2017smart}, modeling air pollution densities \cite{lin2017mining} and visualization of trajectories and crowd flows \cite{bessa2016riobusdata}. Further still, detecting anomalies in urban mobility patterns from physical sensors such as GPS traces and traffic cameras \cite{chawla2012inferring, liu2011discovering}, and CDR \cite{widhalm2015discovering, yin2017generative} is a well-studied topic in the context of optimizing traffic related infrastructure. CDR data have also been used to detect unusual urban events (e.g., elections, emergency events, etc. \cite{dong2015inferring, gundogdu2016countrywide}). Previous works on anomaly detection in transportation have looked at varied aspects of the problem including the detection of and finding the root causes of the anomalies. Pang et al.~\cite{Pang:2011} detect anamolous regions using Likelihood Ratio Tests and, Liu et al.~\cite{liu2011discovering} proposed a formulation for detecting anomalous road blocks using observed minimum distortion and associating causality using frequent subtree mining. We highlight that many of these works rely on private data (obtained directly from transport authorities, or taxicab operators) that is available as stored datasets that allow for studying urban planning related problems as offline case studies. On the contrary, the algorithms we propose here are designed to work with \emph{publicly available}, \emph{real-time} data (available in many metro cities worldwide) which not only useful in offline studies and but also enable the building of practical systems with real-time response times.

There have been several published works on predicting bus arrival times~\cite{zhou2012long, chien2002dynamic, lin1999experimental, bin2006bus} -- we clarify that these works focus on predicting the arrival times (or, $ETA$) of buses \emph{more accurately}, and are complementary to our work in that the algorithms presented in this work consume $ETA$s as input to build trajectories of buses whose identities are not known.

\section{Concluding Remarks}
\label{sec:conc}
In this work, we demonstrate the ability to trace individual bus trajectories from anonymized, \emph{live} ETA records, available in the public domain. We show that the localization of individual buses to downstream bus stops can be performed within reasonable temporal error bounds ($\leq$ 1 min), and provide quantitative  insights on how operational parameters (such as the ETA reporting frequency) affect the localization accuracy. Our evaluations from two major metropolitan cities, Singapore and London, demonstrate that the live, bus-tracking technique presented here can enable applications that were previously only possible with private, fine-grained data (such as smart-fare card transactions).




\small
\section{Acknowledgment}
This material is supported partially by the National Research Foundation, Prime Minister's Office, Singapore under its International Research Centres in Singapore Funding Initiative and under NRF-NSFC Joint Research Grant Call on Data Science (NRF2016NRF-NSFC001-113), and partially by the 
Army Research Laboratory, under agreement number FA5209-17-C-0006. K. Jayarajah's work was supported by an A*STAR Graduate Scholarship. The view and conclusions contained herein are those of the authors and should not be interpreted as necessarily representing the official policies or endorsements, either expressed or implied, of the Army Research Laboratory or the US Government.
\bibliographystyle{IEEEtran}
\bibliography{IEEEabrv,refs}
\end{document}